\documentclass[graybox]{svmult}

\usepackage{type1cm}        
\usepackage{makeidx}         
\usepackage{graphicx}        
\usepackage{multicol}        
\usepackage[bottom]{footmisc}
\usepackage{newtxtext}       %
\usepackage[varvw]{newtxmath}       

\usepackage{import}
\usepackage{booktabs}
\usepackage{tikz} 
\usetikzlibrary{shapes.geometric, arrows}
\usepackage{pgfplots} 
\pgfplotsset{width=\linewidth, compat=1.6} 
\usepackage{pgfplotstable} 

\graphicspath{figures}
\begin{document}

\title*{Towards DNS of Droplet-Jet Collisions of Immiscible Liquids with FS3D}

\author{Johanna Potyka, Jonathan Stober, Jonathan Wurst, Matthias Ibach, Jonas Steigerwald, Bernhard Weigand and Kathrin Schulte}
\authorrunning{J. Potyka, J. Stober, J. Wurst, M. Ibach, J. Steigerwald, B. Weigand and K. Schulte}

\institute{Johanna Potyka and Kathrin Schulte \at Institute of Aerospace Thermodynamics (ITLR), University of Stuttgart, Pfaffenwaldring 31, 70569~Stuttgart, Germany, \email{johanna.potyka@itlr.uni-stuttgart.de} and \email{kathrin.schulte@itlr.uni-stuttgart.de}} 
%
%
\maketitle

\textbf{This is a preprint of the following chapter: J. Potyka, J. Stober, J. Wurst, M. Ibach, J. Steigerwald, B. Weigand and K. Schulte, Towards DNS of Droplet-Jet Collisions of Immiscible Liquids with FS3D, which will be published in the conference proceedings High Performance Computing in Science and Engineering '22, edited by W. E. Nagel, D. H. Kröner and M. M. Resch, Springer reproduced with permission of Springer Nature Switzerland AG. \\
}

\abstract{

In-air microfluidics became a new method for technical production processes with ultra-high throughput formerly performed in micro channels.
Direct Numerical Simulations (DNS) provide a valuable contribution for the fundamental understanding of multiphase flow and later application design. 
This chapter presents a feasibility study with first DNS results of droplet-jet collisions of immiscible liquids using the in-house software Free Surface 3D (FS3D). 
Two cases were investigated with a setup comparable to experiments by Baumgartner et al.~\cite{BAUMGARTNER2020}, where a droplet chain of a glycerol solution hits a jet of silicon oil which encapsulates the droplets. 
The droplets' shapes present are observed to be more complex than comprehensible from the two-dimensional images from the experiments. 
Thus, DNS with FS3D can provide additional information like the surface area or the velocity contributions in order to find analytical models of such collision processes in the future. 
Simulations of such increasingly complex systems require constant improvement of the numerical solver regarding the code's performance. 
Thus, the red-black Gauss-Seidel smoother in the multi-grid solver, the iterative red-black scheme to compute the viscous forces as well as the momentum advection method were enhanced with a cache- and memory usage optimization. An overall performance gain of up to {$33 \%$} was obtained for a representative test case.

}

\section{Introduction}
\label{sec:Introduction}
In-air microfluidics lately became a new alternative to produce fibers, capsules and micro-reactors with ultra-high throughput \cite{Kamperman2018}. Traditionally, such processes are performed on chips with micro channels. The droplet-jet interaction is a chip-free method and thus, not only fast, but also accessible for observation and facilitating further technical curing process such as drying \cite{Kamperman2018} \cite{Visser2018}. The interaction of a droplet with a jet is of interest for technical applications such as capsule- or fiber production for medical or food fabrication purposes. Chen et al.~\cite{CHEN2006-JetDrop} investigated the collision of a water jet with a water droplet of equal diameter, while Kamperman et al. \cite{Kamperman2018} investigated the production of Janus droplets from the collision of two droplet chains or jets with ultra-high throughput. Planchette et al. \cite{PLANCHETTE2018} and Baumgartner et al. \cite{BAUMGARTNER2020} experimentally investigated the structures formed by the interaction of two jets of immiscible liquids with fully wetting liquid combinations and discovered coalescence, which is the merging of the droplet chain into the jet, and separation, which is the rupturing of the jet with capsule formation. A fully wetting liquid with the lower surface tension -in analogy to wetting of surfaces- tends to fully spread on the liquid interface as this is energetically more favorable. 
Those experiments were performed with a jet of silicon oil and a droplet stream of a water-glycerol solution, a combination also used for the simulations in the DNS study on hand. Studying this type of in-air collisions will also broaden the understanding about in-air liquid interaction in general in order to find analytical models.
The modeling of the collision process with respect to predicting the regime boundaries is not yet fully understood \cite{BAUMGARTNER2020,BAUMGARTNER2022}. Baumgartner et. al. \cite{BAUMGARTNER2022} model the stretching process in analogy to the collision of two droplets of equal liquids, but the model still requires some parameters which are found empirically, supported by assumptions transferred from single liquid binary droplet collisions. Part of the assumptions are hard to verify with experimental investigations. Also other regime boundaries like the breakup of the encapsulated liquid inside the jet are not modeled yet. DNS can aid to fill this gap.%

An open research question is for example the development of the interface area of the encapsulated droplets. This information is required for the surface energy evaluation. It is hard to obtain an accurate droplet area development from experiments, as the experiments only provide a two-dimensional view of the process.
Problems like the obstruction of concave shapes are present, but not observable in a two-dimensional view. Three-dimensional simulation results overcome this problem. Additionally, the flow within the oscillating droplets is of interest for obtaining the kinetic energy.
The authors of the present study do not have knowledge of any simulations of droplet-jet collisions of fully-wetting immiscible liquids published until now. Thus, this study aims at proving the feasibility of droplet-jet collision simulations with FS3D by two exemplary droplet-jet interaction cases.
This chapter first presents the relevant mathematical and numerical methods in FS3D necessary for the simulation of immiscible droplet-jet interaction in sec.~\ref{sec:methods}. Two setups are investigated which are extracted from experiments similar to experiments presented by Baumgartner et al. \cite{BAUMGARTNER2020}. The exact setup data of experiments of one coalescence and one separation case were provided by courtesy of D. Baumgartner and C. Planchette from TU Graz. Section~\ref{sec:setup} shows the computational setup employed to reproduce the experiments. Coalescence of the droplet-chain and the jet as well as the separation into encapsulated compound droplets are simulated accordingly and results are shown in sec.~\ref{sec:results}. In a further analysis the encapsulated droplets' surface area as well as the velocity components throughout the collision process are presented. The area and velocity evaluation can be translated into surface energy and kinetic energy of the encapsulated droplet. Future parameter studies will be able to contribute information for analytical modeling approaches for in-air collisions which are hard or impossible to obtain from experiments. %

Recent performance enhancements, which enabled the study of such large and complex interactions in reasonable compute-time, are presented in the second part of this report. An oscillating droplet as a simplification of the encapsulated inner droplet resulting from the collision will serve as a benchmark test case for the performance analysis. The performance enhancements in sec.~\ref{sec:Com_Perf}, along with an efficient three-phase interface reconstruction algorithm by Kromer et al.~\cite{KROMER2021}, enable the computation of droplet-jet interactions with adequate resolutions in reasonable runtime of FS3D on $2048$ cores with $128~\text{MPI-procs./node}$ on HPE Apollo (Hawk).%

\section{Mathematical Description and Numerical Approach}  \label{sec:methods}
The ITLR's in-house CFD code Free Surface 3D (FS3D) performs DNS of incompressible multiphase flows. The code is continuously developed at ITLR since the late 1990s. Several studies show the applicability of FS3D to simulate highly dynamic multiphase processes like droplet deformation \cite{Reutzsch2019a}, droplet impact onto dry and wetted surfaces \cite{FestSantini2021,REN2021}, the atomization of liquid jets \cite{Ertl2017,Ertl2017a} and rivulets \cite{HLRSBericht2021} as well as miscible multi-component interactions \cite{HLRSBERICHT2020}. Such complex fluid interactions require high spatial and temporal resolution. Thus, FS3D is parallelized using Message Passing Interface (MPI) as well as Open Multi-Processing (OpenMP). High parallelization and a high parallel efficiency are required to obtain adequate results within a reasonable time-frame. Current substantial improvements of the overall performance are discussed in sec.~\ref{sec:Com_Perf}.%

FS3D solves the governing equations for mass- and momentum conservation
\begin{equation}
 \label{eq:mass}
  \partial_t \rho + \nabla \cdot \left( \rho \vec{u} \right) = 0 \text{,}	
\end{equation}
\begin{equation}
\label{eq:mom}
  \partial_t \left(  \rho \vec{u}\right)+ \nabla \cdot \left( \rho \vec{u} \otimes \vec{u} \right) = \nabla \cdot \left( \tens{S} - \tens{I} p \right) + \rho \vec{g} + \vec{f}_{\gamma}	
\end{equation}
on finite volumes. In equations (\ref{eq:mass}-\ref{eq:mom}), $\vec{u}$ denotes the velocity vector, $p$ the static pressure, $\vec{g}$ the gravitational acceleration, $\rho$ the density, $t$ denotes the time, $\tens{S}$ is the shear stress tensor and $\tens{I}$ the identity matrix. The volume force $\vec{f}_{\gamma}$ represents surface forces acting at the interface.
The governing equations are solved in a one-field formulation where the different phases are modeled as a single fluid with variable physical properties. FS3D employs the Volume-of-Fluid (VOF) method by Hirt and Nichols \cite{Hirt1981} to identify the different phases. Volume fractions
\begin{equation}
 \label{eq:fdefinition}
  f_m(\textbf{x},t) = \frac{V_m}{V_{\mathrm{cell}}} = \left\{
   \begin{array}{ll}
     0 & \text{outside phase } m,\\
     (0,1) & \text{at an interface of phase } m,\\
     1 & \text{inside phase } m. 
   \end{array}\right.
\end{equation}
are introduced for the disperse phases -in the following the two immiscible liquids- which are advected using the transport equation
\begin{equation}
\label{eq:ftransport}
  \partial_t f_m + \nabla \cdot \left( f_m \mathbf{u} \right) = 0
\end{equation}
for each volume fraction. The corresponding $f_m$-fluxes are calculated with the Piecewise Linear Interface Calculation (PLIC) method by Rider and Kothe \cite{Rider1998} for the interface reconstruction in two phase cells.
A performance optimized version of the three-phase PLIC of Pathak and Raessi \cite{PATHAK2016} is applied in cells where the two immiscible liquids and the surrounding gas are present. 
The cell's density and viscosity are calculated as volume weighted averages. The surface tension forces for the three interfaces present are modeled with the continuous surface stress (CSS) model by Lafaurie~et~al.~\cite{Lafaurie1994} extended to three deformable phases with contact lines employing a superposition approach by Smith~et~al.~\cite{SMITH2002}. 
The general approach for the simulation of three immiscible phases with FS3D was presented by Potyka~et~al.~\cite{SpringerBuchTRR75}. The interface reconstruction within three-phase cells was enhanced by the development of an efficient sequential positioning approach by Kromer~et~al.~\cite{KROMER2022,KROMER2021}. This new method reduced the runtime compared to the previous PLIC implementation in FS3D by $90\%$ which resulted in approximately $20\%$ total runtime reduction of FS3D for representative cases. Additionally, substantial recent improvements of FS3D in general, which are presented in Sec.~\ref{sec:Com_Perf}, enabled the computation of droplet-jet interactions of immiscible liquids within a reasonable runtime.%
\section{Simulation Results}
\subsection{Computational Setup}
\label{sec:setup}
\begin{figure}[tb!]
\centering
 \def\svgwidth{0.55\textwidth}
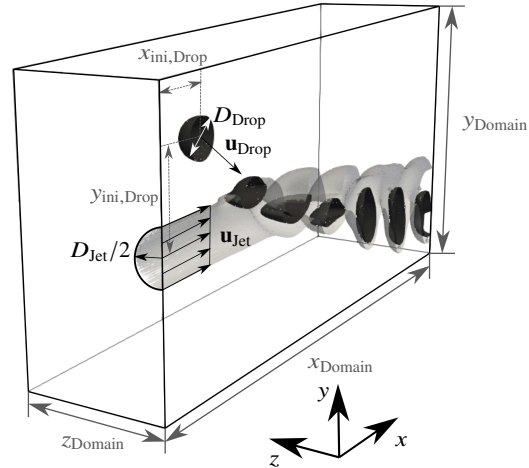
\caption{Visualization of the setup for the droplet-jet interaction. The coordinate origin is in the center of the circle from which the oil jet emerges. It is depicted outside the domain due to visibility. A symmetry plane boundary condition is set in the $x$-$y$-plane along the jet; other boundaries are continuous with an additional inflow of the jet at the $y$-$z$-plane.} \label{fig:DropjetSetup}
\end{figure}

The aforementioned framework within FS3D is employed here to simulate droplet chain-jet interactions. The exact setup data of droplet-jet experiments were provided by courtesy of D. Baumgartner and C. Planchette, the authors of \cite{BAUMGARTNER2020}. Two cases were chosen to obtain a first impression on the further insight DNS can provide: One resulting in coalescence and one exhibits the onset of separation within the domain size simulated. The impact on the surface area development and velocities of the oscillating encapsulated droplet inside the jet were investigated at two different initial jet-perpendicular velocities of the droplets. The setup and exemplary results for the simulations are shown in fig.~\ref{fig:DropjetSetup} for the aqueous droplet chain and the oil jet.
The domain sizes used were $x_\mathrm{Domain} = 3.2~\mathrm{mm}$ for the convergence study and $x_\mathrm{Domain} = 6.4~\mathrm{mm}$ for the comparison of the two cases. The domain size in $y$-direction is in both cases $x_\mathrm{Domain}/2$ and in $z$-direction fixed at $z_\mathrm{Domain} = 0.8~\mathrm{mm}$. The cells are cubes within a Cartesian grid, thus, the resolution is fixed by the resolution in $x$-direction $N_\mathrm{x,Domain}$. A domain decomposition with $64^3$ cells or $48^3$ cells per process, whichever was a reasonable divisor for a chosen resolution, was employed for parallelization. The largest number of cells in $x$-direction for case 2 is $2048$ cells for both domain sizes and thus, roughly $537$ million cells in total. The two largest runs, both for case 2, were computed on $2048$ MPI-processes and ran each for approximately $208.5$ hours. Table~\ref{tab:properties} shows the fluid properties of silicon oil M5 and the $50\%$ glycerol-water solution employed as the two immiscible liquids in the simulations. Table~\ref{tab:cases} lists the initialized parameters of each case. They are visualized in fig.~\ref{fig:DropjetSetup}. The relevant difference of the two cases is the droplet's relative velocity perpendicular to the jet ($y$-direction) as the increased jet velocity is compensated by an equally raised droplet velocity in $x$-direction. All parameters are chosen like in the experiment. The differences between the two cases, other than the relative velocity, are of minor influence on the collision's outcome \cite{BAUMGARTNER2020}.%
\begin{table}[tb!]
\centering
\caption{Liquid and gas viscosities $\mu_m$, densities $\rho_m$ and interfacial tensions $\sigma_{mn}$ of the $50\%$-glycerol-water solution (Droplet), silicon oil M5 (Jet) and air. } \label{tab:properties}
\begin{tabular}{>{\centering}p{0.095\textwidth}>{\centering}p{0.095\textwidth}>{\centering}p{0.14\textwidth}>{\centering}p{0.095\textwidth}>{\centering}p{0.095\textwidth}>{\centering}p{0.095\textwidth}>{\centering}p{0.095\textwidth}>{\centering}p{0.095\textwidth}>{\centering\arraybackslash}p{0.095\textwidth}}
\svhline
$\rho_\mathrm{Drop}$ & $\rho_\mathrm{Jet}$ & $\rho_\mathrm{air}$ & $\mu_\mathrm{Drop}$ & $\mu_\mathrm{Jet}$ & $\mu_\mathrm{air}$ & $\sigma_\mathrm{Drop-air}$ & $\sigma_\mathrm{Jet-air}$ & $\sigma_\mathrm{Drop-Jet}$ \\
$\mathrm{[g / cm^3]}$ & $\mathrm{[g / cm^3]}$ & $\mathrm{[g / cm^3]}$ & $\mathrm{[mPa s]}$ & $\mathrm{[mPa s]}$ & $\mathrm{[mPa s]}$ & $\mathrm{[mN/m]}$ & $\mathrm{[mN/m]}$ & $\mathrm{[mN/m]}$ \\
\hline
$1.1313$ & $0.9134$ & $0.00119$ & $5.24$ & $4.57$ & $0.01824$ & $66.53$ & $19.5$ & $34.3$ \\
\svhline
\end{tabular}
\end{table}
\begin{table}[tb!]
\caption{Initialized parameters visualized in fig.~\ref{fig:DropjetSetup} for both cases. $\Delta T_{\text{Drop}}$: Time between the release of droplets in the chain. Note that the relative motion of the droplet and jet $\Delta u_x$ is more important for the collision dynamics than the absolute values $u_{m,x}$. $u_{\mathrm{Jet,}y}=0$ for both cases. } \label{tab:cases}
\begin{tabular}{p{0.09\textwidth}>{\centering}p{0.09\textwidth}>{\centering}p{0.09\textwidth}>{\centering}p{0.09\textwidth}>{\centering}p{0.09\textwidth}>{\centering}p{0.09\textwidth}>{\centering}p{0.09\textwidth}>{\centering}p{0.18\textwidth}>{\centering\arraybackslash}p{0.09\textwidth}}
\svhline
Case   & $D_{\text{Jet}}$ & $D_{\text{Drop}}$ & $u_{\text{Drop,}x}$ & $u_{\text{Drop,}y}$ & $u_{\mathrm{Jet,}x}$ & $\Delta T_{\text{Drop}}$ & $(x,y)_{\text{ini,Drop}}$ & $\Delta u_x$ \\
       & $\mathrm{[mm]}$ & $\mathrm{[mm]}$   &  $\mathrm{[m/s]}$ & $\mathrm{[m/s]}$ & $\mathrm{[m/s]}$   & $\mathrm{[ms]}$ &  $\mathrm{[mm]}$     & $\mathrm{[m/s]}$ \vspace{0.05cm} \\
\hline
Case 1 & $0.271$            & $0.213$              & $4.12$          & $-3.28$         & $3.74$           & $118$         &  $(0.2,0.6)$ & $0.38$ \\
Case 2 & $0.280$            & $0.230$              & $5.39$          & $-4.29$         & $4.94$           & $121$         & $(0.2,0.6)$ & $0.45$ \\
\svhline
\end{tabular}
\end{table}
In order to obtain the surface area of each encapsulated aqueous droplet inside the jet, a connected component algorithm (CCL) along with the evaluation routines for different quantities was embedded into FS3D to identify and analyze the individual oscillating droplets during runtime \cite{HLRSBericht2021}. For each droplet the center of mass $(x,y,z)_{\mathrm{Drop}}$, as well as the velocity components $u_{\text{Drop},x/y/z}$ were computed and the interface areas $A_\text{Drop}$ summed up. A smooth curve for each of the quantities along the jet was obtained by sampling those quantities at different user-defined time-steps.

\subsection{Results and Discussion}
\label{sec:results}
\begin{figure}[tb!]
\centering
\includegraphics[width =\textwidth]{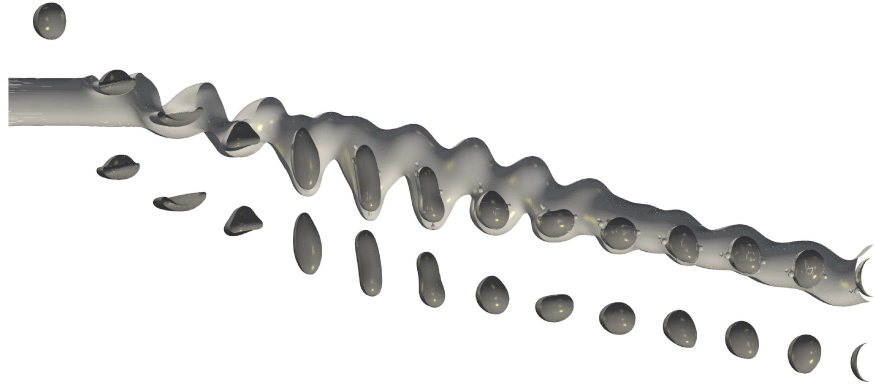} 
\caption{Case 1: Visualization from the symmetry plane under an angle of $25^\circ$ of the coalescence of the full droplet chain and the jet (top) and the oil jet removed (bottom).} \label{fig:case1}
\end{figure}

\begin{figure}[tb!]
\includegraphics[width =\textwidth]{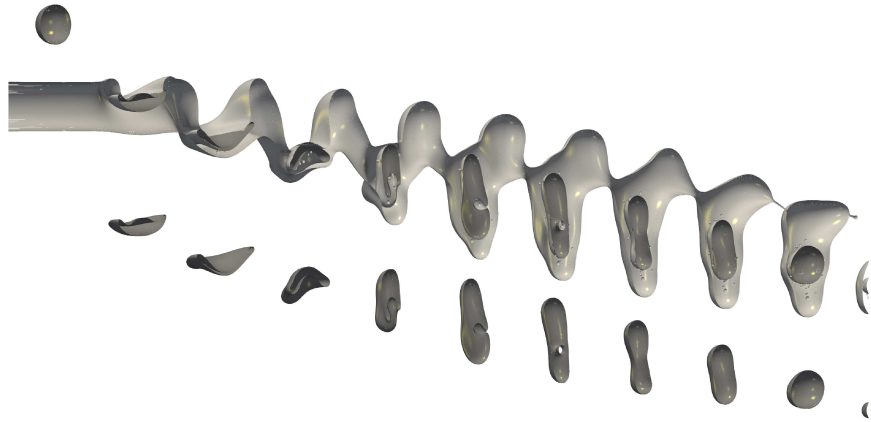} 
\caption{Case 2: Visualization from the symmetry plane under an angle of $25^\circ$ of the onset of separation of the droplet chain and the jet (top). Concave shapes and a hole formation in the oscillating droplet are only visible from the inside and the jet removed (bottom).} \label{fig:case2}
\end{figure}

DNS are able to provide geometry and velocity data in greater detail than experiments can do. In fig.~\ref{fig:case1} and fig.~\ref{fig:case2} a snapshot of the simulated fully developed collision process is shown. A strong deformation in the beginning is followed by an oscillation, which is more complex in case 2 at the higher relative velocity than in case 1. In order to have a clear view of the shapes resulting from the complex oscillation of the inner droplet after the collision, the jet is removed from each picture for comparison. Like in the experiments by Baumgartner et al.~\cite{BAUMGARTNER2020}, case 1 shows a coalescence of the droplet chain with the jet and case 2 exhibits the onset of separation into compound droplets. The droplet shapes present in case 1 are not as complex as in case 2 where the deformation results in concave ``bowl-like'' shapes during the initial disc formation. Further downstream, the droplet in case 2 forms a ``barbell-like" shape and later folds downwards, enveloping some of the outer liquid forming a hole. Such shapes cannot be seen from an outside two-dimensional view in the experiments.%

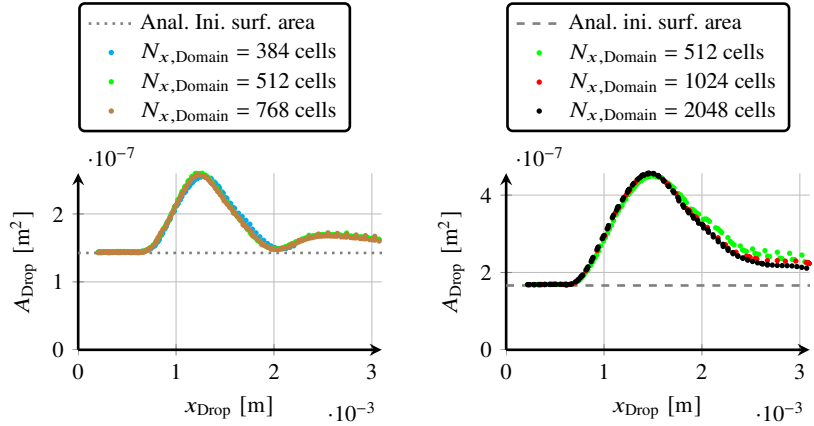
\begin{figure}[tb!]
\centering
\begin{minipage}{0.48\textwidth}
\begin{tikzpicture}
\begin{axis}[
				axis lines=left,
				width=\linewidth,
				height=0.7\linewidth,
				xmin=0.0,xmax=3.1e-3,
            	ymin=0.0,
            	xlabel={$x_\mathrm{Drop}$ $[\mathrm{m}]$},
            	ylabel={$A_\mathrm{Drop}$ $[\mathrm{m^2}]$},
            	grid = major,
		        line width=1pt, 
		       legend style={
					at={(0,1.6)},
					anchor=west,
					legend columns=1,
					cells={anchor=west},
					rounded corners=2pt,}, 
				]
				\addplot[dotted,color=black!50] table[x=x,y=A] {./Txt-files/AnalyticAreaInnerDroplet_A.txt};
				
				\addplot[only marks,mark=o,mark size=0.5 pt,color=cyan] table[x=x,y=A] {./Txt-files/AreaInnerDroplet_A_384cells_SOM5-G50_DropJet_x032cm_2022_04_28.txt};
				
				\addplot[only marks,mark=*,mark size=0.5 pt,color=green] table[x=x,y=A] {./Txt-files/AreaInnerDroplet_2022_04_21_1849_A_SOM5-G50_DropJet_512cells_x032cm.txt};
				
				\addplot[only marks,mark=o,mark size=0.5 pt,color=brown] table[x=x,y=A] {./Txt-files/AreaInnerDroplet_A_768cells_SOM5-G50_DropJet_x032cm_2022_04_28.txt};

\legend{Anal. Ini. surf. area,
        $N_{x,\mathrm{Domain}}=384$ cells,               
        $N_{x,\mathrm{Domain}}=512$ cells,               
        $N_{x,\mathrm{Domain}}=768$ cells,               
        }
\end{axis}
\end{tikzpicture}
\end{minipage}
\begin{minipage}{0.48\textwidth}
\begin{tikzpicture}
\begin{axis}[
				axis lines=left,
				width=\linewidth,
				height=0.7\linewidth,
				xmin=0.0,xmax=3.1e-3,
            	ymin=0.0,
            	xlabel={$x_\mathrm{Drop}$ $[\mathrm{m}]$},
            	ylabel={$A_\mathrm{Drop}$ $[\mathrm{m^2}]$},
            	grid = major,
		        line width=1pt, 
		        legend style={
					at={(0,1.6)},
					anchor=west,
					legend columns=1,
					cells={anchor=west},
					rounded corners=2pt,}, 
				]
				\addplot[dashed,color=black!50] table[x=x,y=A] {./Txt-files/AnalyticAreaInnerDroplet_C.txt};
				
				\addplot[only marks,mark=o,mark size=0.5 pt,color=green] table[x=x,y=A] {./Txt-files/AreaInnerDroplet_C_512cells_SOM5-G50_DropJet_x032cm_2022_04_28_1606.txt};
				\addplot[only marks,mark=*,mark size=0.5 pt,color=red] table[x=x,y=A] {./Txt-files/AreaInnerDroplet_2022_04_28_1606_C_SOM5-G50_DropJet_1024cells_x032cm.txt};
				\addplot[only marks,mark=o,mark size=0.5 pt,color=black] table[x=x,y=A] {./Txt-files/AreaInnerDroplet_C_2048cells_SOM5-G50_DropJet_x032cm_2022_04_28_1606.txt};

\legend{Anal. ini. surf. area,
        $N_{x,\mathrm{Domain}}=512$ cells,               
        $N_{x,\mathrm{Domain}}=1024$ cells,               
        $N_{x,\mathrm{Domain}}=2048$ cells,               
        }
\end{axis}
\end{tikzpicture}
\end{minipage}
\caption{Surface area evolution for case 1 (left) and case 2 (right) at different resolutions of the domain $x_{\mathrm{Domain}}=3.2~\mathrm{mm}$. Case 2 with the higher relative velocity requires a higher resolution.} \label{fig:convergence}
\end{figure}
Before the evolution of the surface area and velocity of the encapsulated droplets are discussed, a convergence study for justification of the chosen resolutions is shown in fig.~\ref{fig:convergence}. The required resolution is dependent on the relative velocity of the droplet and the jet. 
Figure~\ref{fig:convergence} shows the results of the area evaluation at different resolutions, the velocity evolution is not shown, as the differences were minor. The convergence study was performed with a shortened domain ($x_{\mathrm{Domain}} = 3.2~\mathrm{mm}$) as the largest velocities and deformations of the encapsulated droplet's geometry appear shortly after the impact. Thus, computational time can be saved and additionally this allows to increase the resolution beyond the minimum requirements to verify the choice without reaching the scaling limit of FS3D. The largest differences are present in the first retraction phase after the disc formation. A sufficient resolution is reached at $512$ cells for case 1 and $1024$ cells for case 2, resulting in $N_{x,\mathrm{Domain,Case 1}} =1024$ and $N_\mathrm{x,Domain,Case 2} = 2048$ for the large domain.
\begin{figure}[tb!]
\centering
\begin{minipage}[b]{\textwidth}
\begin{tikzpicture}
\begin{axis}[
				axis lines=left,
				height=0.4\linewidth,
				xmin=0.0,xmax=6.3e-3,
            	ymin=0.0,
            	xlabel={$x_{\mathrm{Drop}}$ $[\mathrm{m}]$},
            	ylabel={$A_{\mathrm{Drop}}$ $[\mathrm{m^2}]$},
            	grid = major,
		        line width=1pt, 
		       legend style={
					at={(0.4,0.8)},
					anchor=west,
					legend columns=2,
					cells={anchor=west},
					rounded corners=2pt,}, 
				]
				\addplot[dotted,color=black!50] table[x=x,y=A] {./Txt-files/AnalyticAreaInnerDroplet_A.txt};
				\addplot[dashed,color=black!50] table[x=x,y=A] {./Txt-files/AnalyticAreaInnerDroplet_C.txt};
				
				\addplot[only marks,mark=o,mark size=0.5 pt,color=green] table[x=x,y=A] {./Txt-files/AreaInnerDroplet_2022_04_28_1606_A_SOM5-G50_DropJet_1024cells_x064cm.txt};
				\addplot[only marks,mark=o,mark size=0.5 pt,color=green!40!black!60!] table[x=x,y=A] {./Txt-files/AreaInnerDropletFirst_2022_04_28_1606_A_SOM5-G50_DropJet_1024cells_x064cm.txt};

				\addplot[only marks,mark=o,mark size=0.5 pt,color=red] table[x=x,y=A] {./Txt-files/AreaInnerDroplet_C_2048cells_SOM5-G50_DropJet_x064cm_2022_04_28_1606.txt};
				\addplot[only marks,mark=o,mark size=0.5 pt,color=red!40!black!60!] table[x=x,y=A] {./Txt-files/AreaInnerDropletFirst_C_2048cells_SOM5-G50_DropJet_x064cm_2022_04_28_1606.txt};

\legend{Ini. surf. area Case 1,
        Ini. surf. area Case 2,
        Case 1,
        First droplet  Case 1,
        Case 2,
        First droplet  Case 2,
        }
\end{axis}
\end{tikzpicture}

\end{minipage}

\begin{minipage}[b]{\textwidth}
\begin{tikzpicture}
\begin{axis}[
				axis lines=left,
				height=0.4\linewidth,
				xmin=0.0,xmax=6.3e-3,
            	ymin=-1.0,ymax=7.5,
            	xlabel={$x_\mathrm{Drop}$ $[\mathrm{m}]$},
            	ylabel={$u_{\mathrm{Drop}}$ $[\mathrm{m/s}]$},
            	grid = major,
		        line width=1pt, 
		       legend style={
					at={(0.3,1.1)},
					anchor=west,
					legend columns=2,
					cells={anchor=west},
					rounded corners=2pt,}, 
				]
				\addplot[only marks,mark=o,mark size=0.5 pt,color=green] table[x=x,y=ures] {./Txt-files/u_res_InnerDroplet_2022_04_28_1606_A_SOM5-G50_DropJet_1024cells_x064cm.txt};
				\addplot[only marks,mark=o,mark size=0.5 pt,color=green!20!black!60!] table[x=x,y=ures] {./Txt-files/u_res_InnerDropletFirst_2022_04_28_1606_A_SOM5-G50_DropJet_1024cells_x064cm.txt};

				\addplot[only marks,mark=o,mark size=0.5 pt,color=green!60!cyan!40!] table[x=x,y=uz] {./Txt-files/u_z_InnerDroplet_2022_04_28_1606_A_SOM5-G50_DropJet_1024cells_x064cm.txt};
				\addplot[only marks,mark=o,mark size=0.5 pt,color=green!20!black!60!] table[x=x,y=uz] {./Txt-files/u_z_InnerDropletFirst_2022_04_28_1606_A_SOM5-G50_DropJet_1024cells_x064cm.txt};

				\addplot[only marks,mark=o,mark size=0.5 pt,color=red] table[x=x,y=ures] {./Txt-files/u_res_InnerDroplet_C_2048cells_SOM5-G50_DropJet_x064cm_2022_04_28_1606.txt};
				\addplot[only marks,mark=o,mark size=0.5 pt,color=red!60!black!60!] table[x=x,y=ures] {./Txt-files/u_res_InnerDropletFirst_C_2048cells_SOM5-G50_DropJet_x064cm_2022_04_28_1606.txt};
				
				\addplot[only marks,mark=o,mark size=0.5 pt,color=red!60!magenta!40!] table[x=x,y=uz] {./Txt-files/u_z_InnerDroplet_C_2048cells_SOM5-G50_DropJet_x064cm_2022_04_28_1606.txt};
				\addplot[only marks,mark=o,mark size=0.5 pt,color=red!60!black!60!] table[x=x,y=uz] {./Txt-files/u_z_InnerDropletFirst_C_2048cells_SOM5-G50_DropJet_x064cm_2022_04_28_1606.txt};

\legend{$u_{\mathrm{res}} = \sqrt{u_x^2 + u_y^2}$ Case 1,
        $u_\mathrm{res}$ Case 1 First droplet,
        $u_z$ of the half droplet Case 1,
        $u_z$ Case 1 First droplet,
        $u_{\mathrm{res}} = \sqrt{u_x^2 + u_y^2}$ Case 2,
        $u_\mathrm{res}$ Case 2 First droplet,
        $u_z$ of the half droplet Case 2,
        $u_z$ Case 2 First droplet,
        }
\end{axis}
\end{tikzpicture}
\end{minipage}
\caption{Surface area (top) and velocity (bottom) of the encapsulated oscillating droplet for case 1 and 2  with $x_\mathrm{Domain}=6.4~\mathrm{mm}$ ($N_{x,\mathrm{Domain}}$ case 1: $1024$ cells, case 2: $2048$ cells). The spherical surface of the initialized droplet is shown for reference. The area and velocity course of the first droplet to hit the jet is marked and differs slightly. The resulting velocity of the droplet cut at the symmetry plane in $z$-direction indicates the duration of spreading- and retracting phases.} \label{fig:area_velocity}
\end{figure}
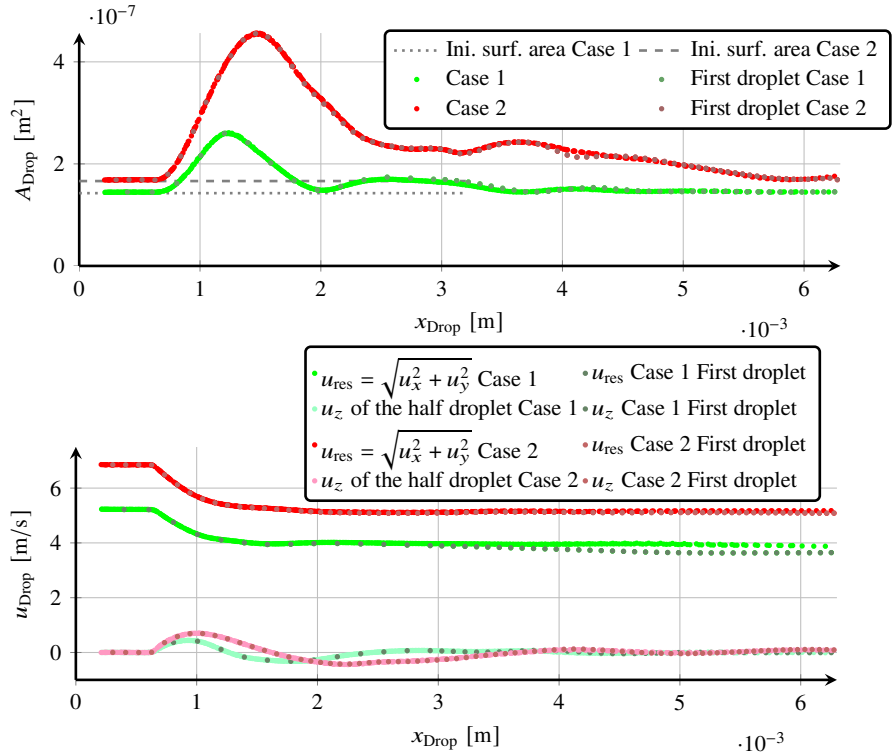
The surface area and velocity of the inner droplet were evaluated for the two test cases with the large domain size of $x_{\mathrm{Domain}} = 6.4~\mathrm{mm}$ in order to capture the relevant part of the collision dynamics. The larger the relative velocity perpendicular to the jet, the thinner the spreading disc of the droplet on the jet becomes. A higher velocity setting leads to a much higher maximum in the surface area which is also shifted further downstream as the spreading takes more time. The amplitude of the deformation is much larger which deforms the jet until separation of the oil jet takes place. With the complex and concave shapes involved, an estimation of the surface area of the inner droplet from photographs of experimental results will underestimate the surface area. With the advantage of highly resolved DNS at hand, this problem vanishes. Figure~\ref{fig:area_velocity} shows the area and velocity contributions of the encapsulated droplet along the jet. It is visible, that the droplet is decelerated until a new constant velocity is reached inside the jet. The first droplet behaves slightly different from all subsequent droplets: In case 1 the first droplet is decelerated slightly more at the later collision stages, but this does not have much effect on the area evolution. On the other hand in case 2 the area development differs during the oscillation, which is most pronounced after the center of mass of the droplet reaches $x_\mathrm{Drop} = 4~\mathrm{mm}$. The velocity development is not affected much in case 2 with the high initial inertia. The second droplet which hits the jet already exhibits the same area and velocity course like all subsequent droplets in both cases. Thus, the collision complex can be considered quasi-stationary, as soon as the first droplet leaves the domain. This is advantageous as the simulation can be stopped as soon as the second droplet touches the domain boundary. In the fully developed situation, the velocity of the half droplet in $z$-direction reaching zero resulting velocity the first time after the collision does not coincide with the maximum surface area. The surface area maximum appears slightly earlier. The velocity of half of the droplet cut in the symmetry plane is also shown, as the velocity of the full droplet is zero due to symmetry. This half droplet's velocity indicates additionally to the depiction of the droplet shapes in figure~\ref{fig:case1} and figure~\ref{fig:case2}, that modeling the collision as a simple disc expansion and retraction or an oscillating ellipsoid is not sufficient. More complex movement is involved due to the liquid's interaction, especially in case~2 with a higher relative velocity perpendicular to the jet. 
In both cases the velocity of the droplet parallel to the original jet direction does not change much throughout the collision. The velocity in $y$-direction orthogonal to the original jet direction on the other hand is decreasing by a large factor as kinetic energy is transferred into surface- and dissipated energy. %

This first analysis clearly shows the advantages of highly resolved temporal and spatial data on in-air microfluidics. It also proves the feasibility to obtain and evaluate such results with our in-house multiphase simulation code FS3D when supercomputing capacities on HPE Apollo (Hawk) are employed.

\section{Computational Performance} \label{sec:Com_Perf}
\begin{figure}[tb!]
	\centering
	\includegraphics[width=0.8\textwidth]{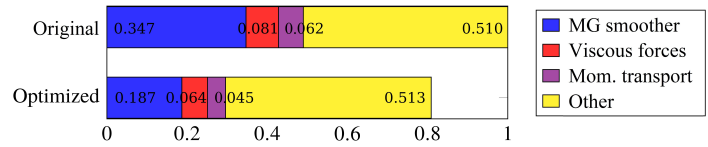}
	\caption{Fractions of computational time of the three most time-consuming routines for an exemplary simulation of an oscillating droplet with a computational grid consisting of $512^3$ cells with $512$ MPI-processes using the original solver (top) and the fully optimized solver (bottom).}
	\label{fig:clogain}
\end{figure}

An efficient run of FS3D on Hawk is desirable in order to reach an economical use of resources along with a realistic runtime for the computation of large systems. Constant improvements of FS3D's performance are required in order to keep up with supercomputing hardware developments. For this reason, we further improved the parallel performance of FS3D by optimizing the cache usage within the three most time-consuming routines during a solution cycle. These routines are first, the red-black Gauss-Seidel smoother routine of the employed multigrid (MG) solver for the pressure Poisson equation, referred to as MG smoother in the following, second, the implicit scheme for the calculation of the viscous forces and third, the routine for the calculation of the momentum transport. The upper part of fig.~\ref{fig:clogain} shows the fraction of the computational time of these three routines with respect to the total time of an average single solution cycle for an exemplary simulation, which was performed with $512$ MPI-processes with the original solver. It is visible, that the solver spends approximately $50\%$ of the time within these three routines. This percentage of time can, however, be drastically reduced by optimizing the cache reuse (see lower part of fig.~\ref{fig:clogain}). As fig.~\ref{fig:clogain} was obtained from measurements with the performance analysis tool Maqao, the difference in the measured performance compared to the following measurements without such a tool may originate from overhead and additional cache usage produced by this tool. A description of the applied optimization as well as a detailed analysis of its effects on the overall parallel performance of FS3D is presented in the following. 

\subsection{Cache-Line Optimization}  
\begin{figure}[tb!]
	\centering
	\includegraphics[width=0.7\textwidth]{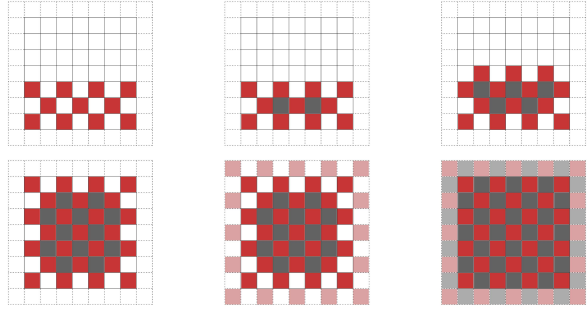}
	\caption{Cache-line optimized red-black iteration scheme. After three slices of updating \textit{red} cells, the \textit{black} cells are processed always one slice behind the \textit{red} iteration.}\label{fig:redblackclo}
\end{figure}

The idea behind cache-line optimization is to reduce the number of cache misses throughout a calculation step to avoid loading data from main memory which leads inevitably to a loss of performance. We applied a blockwise sweeping cache-line optimization technique \cite{Wauligmann2021} to the MG smoother and to the implicit scheme for the calculation of the viscous forces. Both employ a red-black Gauss Seidel iteration scheme. The scheme uses a 7-point stencil in which 6 adjacent \textit{red} cells are used to update a single \textit{black} cell and vice versa. In the original and previously used implementation of the red-black scheme, the solution algorithm sweeps first through all \textit{red} and subsequently through all \textit{black} cells. After each sweep through either \textit{red} or \textit{black} cells, an MPI halo exchange is performed to update the corresponding cells. Even though the implementation of this widely used algorithm is straight forward, it has the disadvantage that the data of the domain, which is swept through, is in general too large to fit into the caches during the computation. This leads to a high number of cache misses and consequently to a significant reduction of the speed of data access, as it has to be loaded twice from main memory. \\
The applied cache-line optimization uses data of the \textit{red} cells that is already in the cache now for updating the \textit{black} cells. Such an approach can be achieved by a blockwise instead of a domainwise sweeping through all cells, which is exemplary visualized in fig.~\ref{fig:redblackclo} for a two-dimensional data set. As soon as the iteration has progressed through three slices of data for the \textit{red} cells, the algorithm starts also to update the data of \textit{black} cells in an alternating way, staying always one slice behind the \textit{red} update sweep (see first row in fig.~\ref{fig:redblackclo}). Due to that, the inner \textit{black} cells are computed without an MPI halo exchange and only \textit{red} cell data is used, which is already in the cache. The MPI halo exchange and outer black cells are updated later on. The blockwise sweeping optimizes the spatial locality of the data within the cache, which results in almost doubled execution speed of the routine. \\
In contrast to the iterative routines, the performance of the routine which deals with the momentum transport was optimized in a different way by a clever reorganization and nesting of previously separated loops, which also increased the spatial locality within the caches. The overall effect of the applied cache-line optimization on all routines was analyzed via strong and weak scaling measurements, which will be presented in the next section.

\subsection{Benchmark Case and Performance Analysis} 
As in previous studies, we use a single isothermal oscillating water droplet in ambient air as a representative benchmark case in order to analyze the computational performance of FS3D \cite{HLRSBericht2021,HLRSBERICHT2020}. A droplet with an equivalent spherical diameter of $2~\mathrm{mm}$ is initialized at the center within a cubic domain with a length of $8~\mathrm{mm}$ per edge. The oscillation is provoked by initializing the droplet as an ellipsoid with an oblate shape with semi-principle axis of $a=b=1.357~\mathrm{mm}$ and $c=0.543~\mathrm{mm}$. The physical properties of the water and the ambient air correspond to standard ambient conditions $T=293.15~K$ and pressure of one atmosphere. \newline 
A strong and weak scaling measurement evaluates the parallel performance and scalability of FS3D by using the amount of completed calculation cycles per hour (CPH). The CPH are estimated as an average over 10 simulations each with a fixed walltime of 30 minutes and by subtracting the exact time of initialization afterwards. The number of processors which corresponds to the number of MPI-processes is varied from $1$ to $16^3$. Hyperthreading or hybrid parallelization on a loop level with OpenMP was not employed here. All simulations with $\ge 128$ MPI-processes employ all $128$ processors per node. Furthermore, no additional data was written out during the simulations to enable a performance analysis which is not influenced by the interconnected file system. The source code was compiled with the gfortran-Compiler using the Ofast optimization option and an additional Link-Time Optimization (LTO). This setting leads to the best performance of FS3D as identified by Schlottke et al.~\cite{HLRSBericht2021}.

\subsubsection{Strong Scaling} 
\label{sec:Strong_Scaling}
\begin{figure}[tb!]
	\centering
	\includegraphics[width=0.7\textwidth]{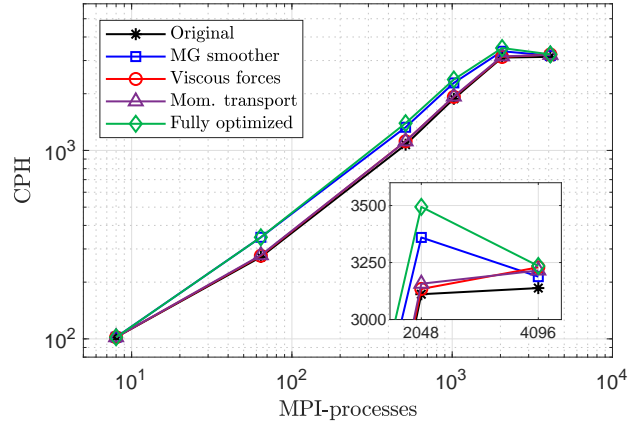}
	\caption{Calculation cycles per hour (CPH) over the number of MPI-processes of the strong scaling analysis for the original solver, solver with individual optimized routines and fully optimized solver.}
\label{figure:strong_scaling}
\end{figure}

The strong scaling performance was performed with a baseline case with a spatial resolution of $512^3$ grid cells while the number of MPI-processes was successively increased from $2^3$ up to $16^3$. The setups of all simulated cases and the achieved CPH are summarized in table~\ref{table:sssetup}. The table also shows the percentage increase of solved cycles with respect to the original solver. The result of the calculation by using a single core is not considered here as no single calculation cycle was completed within the allotted walltime. Note that the local computational domains of simulations for which $2 \times 8^3$ and $4 \times 8^3$ MPI-processes are employed are not cubic. 
\begin{table}[tb!]
\caption{Setup for strong scaling and calculation cycles per hour (CPH) for the original solver, the original solver with individual optimized routines and for the fully optimized solver.}
\label{table:sssetup}
\centering
\begin{tabular}{p{2.8cm}p{1.3cm}p{1.3cm}p{1.3cm}p{1.3cm}p{1.3cm}p{1.3cm}}
\svhline\noalign{\smallskip}
Problem size & \multicolumn{6}{c}{$512^3$}\\
\noalign{\smallskip}\hline\noalign{\smallskip}
MPI-processes     & $2^3$   &$4^3$   &$8^3$  &  $2 \times 8^3$  &  $4 \times 8^3$  & $16^3$ \\
Cells per process & $256^3$ &$128^3$ &$64^3$ & $64^2 \times 32$ & $64 \times 32^2$ & $32^3$ \\
Nodes & $1$ &$1$ &$4$ & $8$ & $16$ & $32$ \\
\noalign{\smallskip}\svhline\noalign{\smallskip}
CPH (Original) & $102$ & $272$ & $1073$ & $1881$ & $3111$ & $3138$ \\
 \noalign{\smallskip}\hline\noalign{\smallskip}
CPH (MG smoother) & $102$ & $346$    & $1327$    & $2282$    & $3360$    & $3188$ \\
                  & $-$  & $+27.0\%$ & $+23.7\%$ & $+21.3\%$ & $+8.00\%$ & $+1.57\%$\\
CPH (Viscous forces) & $102$ & $276$     & $1118$    & $1924$    & $3134$    & $3229$ \\
                     & $-$   & $+1.40\%$ & $+4.12\%$ & $+2.27\%$ & $+0.75\%$ & $+2.87\%$\\ 
CPH (Mom. transport) & $102$ & $278$ & $1117$ & $1932$ & $3157$  & $3215$ \\
 & $-$ & $+2.13\%$ & $+4.10\%$ & $+2.71\%$ & $+1.49\%$ & $+2.43\%$\\
 \noalign{\smallskip}\hline\noalign{\smallskip}
CPH (Fully optimized) & $102$ & $346$ & $1395$ & $2379$ & $3494$ & $3235$ \\
 & $-$ & $+27.1\%$ & $+29.9\%$ & $+26.5\%$ & $+12.3\%$ & $+3.07\%$\\
\noalign{\smallskip}\svhline\noalign{\smallskip} 
\end{tabular}
\end{table}
As can be seen, optimizing the MG smoother leads to a significant increase of the overall performance of up to $27\%$ for the investigated cases. The increase stays above $20\%$ for local domains with more than $64^2 \times 32$ grid cells but reduces when more than $1024$ MPI-processes are used, which already lead to a better cache reuse due to smaller array sizes. The optimized calculation of the momentum transport and the viscous forces leads also to an increase of the overall performance, which ranges approximately between $1-4\%$. Even though this achieved improvement percentage is only single-digit, the increased performance of these routines is nevertheless still remarkable, if one takes into account the lower fraction of compute time compared to the MG smoother the solver spends within these routines during a solution cycle. This can be seen in the lower part of fig.~\ref{fig:clogain}, which shows the normalized computational time of the three optimized routines with respect to the time of a whole original as well as optimized solution cycle for an exemplary case. The applied cache-line optimization leads to an enormous decrease of the computational time of the MG smoother by $46\%$, of the calculation of the viscous forces by $21\%$ and of the momentum transport by $27\%$ in this representative case. This leads to a total reduction of the computational time during a solution cycle by $19\%$ for the present case with $512^3$ grid cells and $512$ MPI-processes. As the MG smoother is the most time consuming routine during a solution cycle, the overall improvement is mainly due to the optimized MG smoother, which is also visualized in fig.~\ref{figure:strong_scaling}. The fully optimized code solves now up to almost $30\%$ more CPH than the original solver.
\subsubsection{Weak Scaling} 
\label{sec:Weak_Scaling}
\begin{figure}[tb!]
	\centering
	\includegraphics[width=0.7\textwidth]{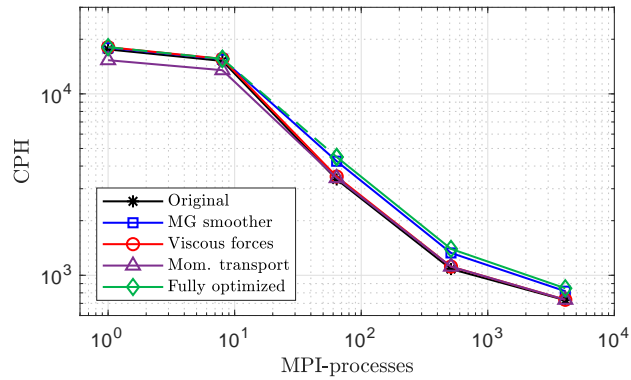}
	\caption{Calculation cycles per hour (CPH) over the number of MPI-processes of the weak scaling analysis for the original solver, solver with individual optimized routines and fully optimized solver.}
\label{figure:weak_scaling}
\end{figure}

The analysis of the weak scaling keeps the amount of grid cells per MPI-process constant at $64^3$ and increases the number of MPI-processes successively from $1$ to $16^3$ which corresponds to global domain sizes between $64^3$ and $1024^3$ grid cells. The corresponding setups and the achieved CPH are summarized in table~\ref{table:wssetup}. The use of the optimized MG smoother increased the CPH significantly if more than $8$ MPI-processes are employed. The greatest increase was observed for $64$ MPI-processes, where the CPH increased by more than $26\%$. The percentage improvement decreases only slightly with increasing processor numbers and is still above $12\%$ when $4096$ MPI-processes are employed. For simulations with less than $8$ MPI-processes, the optimized cache usage in the MG smoother routine seems, however, to have only a small effect on the performance as the CPH increases only by $2-3\%$. The optimized calculation of the viscous forces improves the overall performance by $2.6-3.7\%$ over a wide range of MPI-process numbers. The optimized momentum transport even lead to a strong decrease of the estimated CPH at low MPI-process numbers (see fig.~\ref{figure:weak_scaling}). However, for simulations with more than $8$ MPI-processes, which is always the case for production runs and the cache-line optimization was intended for, the overall improvement of the parallel efficiency is again significant. For the fully optimized solver, the calculated CPH increase by $33\%$ for $64$ MPI-processes and by $16\%$ for $4096$ MPI-processes, demonstrating the huge benefit when cache misses are reduced.
\begin{table}[tb!]
\caption{Setup for weak scaling and calculation cycles per hour (CPH) for the original solver, the original solver with individual optimized routines and for the fully optimized solver.}
	\label{table:wssetup}
\centering
\begin{tabular}{p{2.8cm}p{1.3cm}p{1.3cm}p{1.3cm}p{1.3cm}p{1.3cm}p{1.3cm}}
\hline\noalign{\smallskip}
Cells per process & \multicolumn{5}{c}{$64^3$}\\
\noalign{\smallskip}\hline\noalign{\smallskip}
Problem size & $64^3$ & $128^3$ & $256^3$ & $512^3$ & $1024^3$  \\
MPI-processes & $1$ & $2^3$   &$4^3$   &$8^3$  & $16^3$ \\
Nodes & $1$ &$1$ &$1$ & $4$ & $32$ \\
\noalign{\smallskip}\svhline\noalign{\smallskip}
CPH (Original) & $17548$  & $15188$ & $3375$ & $1083$ & $729$ \\
\noalign{\smallskip}\hline\noalign{\smallskip}
CPH (MG smoother) & $17962$ & $15522$ & $4254$ & $1328$ & $817$ \\
 & $+2.15\%$ & $+2.20\%$ & $+26.1\%$ & $+22.6\%$ & $+12.0\%$ \\
CPH (Viscous forces) & $18096$ & $15587$ & $3501$ & $1117$ & $733$ \\
 & $+2.91\%$ & $+2.63\%$ & $+3.73\%$ & $+3.09\%$ & $+0.45\%$\\ 
CPH (Mom. transport) & $15336$ & $13504$ & $3433$ & $1116$ & $733$ \\
 & $-12.8\%$ & $-11.1\%$ & $+1.72\%$ & $+3.09\%$ & $+0.45\%$\\
 \noalign{\smallskip}\hline\noalign{\smallskip}
CPH (Fully optimized) & $15840$ & $14006$ & $4486$ & $1398$ & $848$ \\
 & $-9.92\%$ & $-7.78\%$ & $+32.9\%$ & $+29.1\%$ & $+16.3\%$\\
\noalign{\smallskip}\svhline\noalign{\smallskip} 
\end{tabular}
\end{table}

\section{Conclusion}
Droplet-jet interactions in air were computed with the multiphase DNS code FS3D. The simulations allow a view at the shape of the droplets, which cannot be extracted from a two-dimensional plane view in experiments. The shapes are complex and three dimensional, especially at a higher relative impact velocity of the droplets. The surface area of the inner droplet as well as the velocity was evaluated for two representative cases at different relative jet-perpendicular velocities. The required minimum resolution was evaluated for each case. It was found that a quasi-stationary solution can be assumed already for the second droplet to hit the jet. The results clearly show that only supercomputers are able to simulate droplet-jet interactions with adequate resolutions in reasonable time. FS3D's modeling approaches and internal routines for the identification of single droplets, their surface area as well as their velocities were tested successfully. With this, FS3D provides a framework for immiscible liquid interaction studies. Simulations with a range of parameters relevant in droplet-jet interactions are feasible, but with high demands of computational resources. The recent improvements of the performance with cache optimization yielded an approximately $30\%$ reduction of the overall runtime for representative cases. Considering the usage of roughly $30$ million Core-h with FS3D on Hawk within the last year, the total amount of reduced computational time is considerable. Summarized, it is feasible to simulate droplet-jet collisions with the current methods in FS3D on the supercomputer Hawk within a reasonable time frame now. This clears the path for detailed evaluations of such processes, which can eventually lead to analytical modeling approaches of microfluidics in air.%

%


\newenvironment{acknowledgments}%
{\null\begin{center}%
 	\bfseries Acknowledgments\end{center}}%
{\null}
\begin{acknowledgments}
The authors kindly acknowledge the High Performance Computing Center Stuttgart (HLRS) for support and supply of computational resources on the HPE Apollo (Hawk) platform under the Grant No. FS3D/11142. The authors also gratefully acknowledge the financial support from the Deutsche Forschungsgemeinschaft (DFG, German Research Foundation) through  the projects SFB-TRR75 (84292822), DROPIT/GRK 2160/2 and EXC 2075 (390740016). We also acknowledge the financial support of the Friedrich and Elisabeth Boysen Foundation under grant BOY-160. \\
We want to thank C. Planchette and D. Baumgartner from TU Graz for the provision of experimental reference data and the fruitful discussions on droplet-jet interactions and Moritz Heinemann from VISUS for the three-phase PLIC plugin for Paraview (https://github.com/UniStuttgart-VISUS/tpf).
\end{acknowledgments}

\bibliographystyle{spmpsci}
\bibliography{references}

\eject
\end{document}